\documentclass {elsart}
\usepackage{graphicx}

\usepackage{graphics}
\usepackage{graphicx}
\usepackage{epsfig}

\usepackage{amssymb}
\usepackage{amsmath}

\begin{document}

\begin{frontmatter}



\title{Generalized Statistics Variational Perturbation Approximation using q-Deformed Calculus}


\author[SRC]{R. C. Venkatesan\corauthref{cor}}
\corauth[cor]{Corresponding author.}
\ead{ravi@systemsresearchcorp.com}
\author[UNLP]{A. Plastino}
\ead{plastino@venus.fisica.unlp.edu.ar}

\address[SRC]{Systems Research Corporation,
Aundh, Pune 411007, India}
\address[UNLP]{IFLP, National University La Plata \&
National Research Council (CONICET)\\ C. C., 727 1900, La Plata,
Argentina}

\begin{abstract}
A principled framework to generalize variational perturbation
approximations (VPA's) formulated within the ambit of the
nonadditive statistics of Tsallis statistics, is introduced. This is
accomplished by operating on the terms constituting the perturbation
expansion of the generalized free energy (GFE) with a variational
procedure formulated using \emph{q-deformed calculus}. A candidate
\textit{q-deformed} generalized VPA (GVPA) is derived with the aid
of the Hellmann-Feynman theorem. The generalized Bogoliubov
inequality for the approximate GFE are derived for the case of
canonical probability densities that maximize the Tsallis entropy.
Numerical examples demonstrating the application of the
\textit{q-deformed} GVPA are presented. The qualitative distinctions
between the \textit{q-deformed} GVPA model \textit{vis-\'{a}-vis} prior GVPA
models are highlighted.
\end{abstract}

\end{frontmatter}

\section{Introduction}
The generalized (nonadditive) statistics of Tsallis' [1,2] has
recently been the focus of much attention in statistical physics,
and allied disciplines.  Nonadditive statistics\footnote{The terms
generalized statistics, nonadditive statistics, and nonextensive
statistics are used interchangeably.}, which generalizes the
extensive Boltzmann-Gibbs-Shannon (B-G-S) statistics, has much
utility in a wide spectrum of disciplines ranging from complex
systems and condensed matter physics to financial mathematics
\footnote{A continually updated bibliography of works related to
nonextensive statistics may be found at
http://tsallis.cat.cbpf.br/biblio.htm.}. Recent works have extended
the scope of Tsallis statistics, by demonstrating its efficacy in
lossy data compression in communication theory [3] and machine
learning [4].

Variational perturbation approximations (VPA's) [5] are extensively
employed in quantum mechanics and statistical physics [6]. The first
attempt to generalize VPA's to the case of Tsallis statistics was
performed by Plastino and Tsallis [7]. This proof-of-principle
analysis for generalized VPA (GVPA) models [7] established the
predominance of concavity of the measure of uncertainty over
extensivity (as defined within the context of B-G-S statistics).
Further work by Lenzi, Malacarne, and Mendes [8] and Mendes
\textit{et. al.} [9] demonstrated the workings of a GVPA model for
the GFE expanded to include second-order terms, using a classical
harmonic oscillator as an example. More recently, Lu, Cai, and Kim
[10] demonstrated that the inclusion of higher-order terms can
significantly improve the results of a GVPA.

The generic procedure for GVPA's is as follows: $ (i) $ evaluation
the canonical probability distribution $ p_n $ that maximizes the
Tsallis entropy, and, formulation of the generalized free energy (GFE) [11]
\begin{equation}
F_q = U_q  - \frac{1}{\beta }S_q =  - \frac{1}{\beta }\frac{{\tilde
Z^{1 - q}  - 1}}{{1 - q}} =  - \frac{1}{\beta }\ln _q \tilde Z ,
\end{equation}
where, $ S_q $ is the Tsallis entropy (defined in Section 2), $ U_q
$ is the generalized internal energy, $ \beta $ (the energy Lagrange multiplier) is the "inverse thermodynamic temperature"\footnote{Note that for constraints expressed in the form of normal averages, $\beta$ is replaced by $\tilde\beta=\beta/q$ as is discussed in Section 3.2 and Appendix A of this paper.}, and $ \tilde Z $ is canonical partition
function, $ (ii) $ assuming a Hamiltonian of the system as
\begin{equation}
H = H_0  + \lambda H_1,
\end{equation}
where, $H_0$ is a Hamiltonian of a soluble system and $ \lambda H_1
$ is a perturbation to $H_0$.  Here, $ \lambda \in[0,1] $ is a
perturbation parameter, $ (iii) $ perturbation expansion of the GFE
\begin{equation}
\begin{array}{l}
F_q(\lambda)  = F_q^{(0)}  + \left. {\lambda \delta ^{\left( 1
\right)} F_q } \right|_{\lambda  = 0}  + \left. {\frac{{\lambda ^2
}}{{2!}}\delta ^{\left( 2 \right)} F_q } \right|_{\lambda  = 0}  +
...........,
\end{array}
\end{equation}
where: $ \delta ^{\left( k \right)} F_q  = \frac{{d^k F_q
}}{{d\lambda ^k }}; k=1,... $, and, $ (iv) $ solving (2) with the
aid of the Hellmann-Feynman theorem [12]
\begin{equation}
\frac{{dE_n }}{{d\lambda }} = \left\langle n
\right|\frac{{dH}}{{d\lambda }}\left| n \right\rangle  =
\left\langle n \right|H_1 \left| n \right\rangle, \\
\end{equation}
where, $ E_n $ is the $ n^{th} $ energy eigenvalue of the
Hamiltonian $ H $, and, $ (v) $ evaluating the generalized
Bogoliubov inequality [7-10]

\begin{equation}
F_q  \le F_q^{(0)}  + \left\langle {H_1 } \right\rangle^{(0)},
\end{equation}
where, $  \left\langle {H_1 } \right\rangle^{(0)} $ is the
generalized expectation of $ H_1 $ in (2) evaluated with $ \lambda=0
$.  Note that while the proof-of-principle analysis in [7]
implicitly assumed the above steps $ (i)-(v) $  to have been
performed, their explicit implementation was demonstrated in [8-10].
Further,the expression for the GFE (1) has recently been the object
of much research and debate.  Most generally, the inverse temperature
$ \beta $, which in essence is the Lagrange multiplier associated
with the internal energy relates to the thermodynamic temperature $
T $ as: $ \beta  = \frac{1}{{k_BT }} $, where $k_b$ is the Boltzmann constant (sometimes set to unity for the sake of convenience) only in the limiting case
$q\rightarrow 1$. Prominent attempts to mitigate this issue are
those by Abe \textit{et. al.} [13], Abe [14], amongst others.

 Generalized statistics
and the problem of obtaining maximum Tsallis entropy canonical
probability distributions has been associated with many forms of
constraints. These include the linear constraints originally
employed by Tsallis [1] (also known as normal averages) of the form:
$ \left\langle A \right\rangle = \sum\limits_i {p_i } A_i $, the
Curado-Tsallis (C-T) constraints [15] of the form:  $ \left\langle A
\right\rangle _q  = \sum\limits_i {p_i^q } A_i  \ $, and, the
normalized Tsallis-Mendes-Plastino (T-M-P) constraints [11](also
known as $q$-averages) of the form:  $ \left\langle {\left\langle A
\right\rangle } \right\rangle _q  = \sum\limits_i {\frac{{p_i^q
}}{{\sum\limits_i {p_i^q } }}A_i } \ $. The normalized T-M-P
constraints render the canonical probability distribution to be
\textit{self-referential}, owing to the dependence of the
expectation value on the normalized \textit{pdf}. A fourth form of
constraints are the optimal Lagrange multiplier (OLM) constraints
[16, 17].  A work by Ferri, Martinez, and Plastino [18] introduces a
methodology to "rescue" the normal averages constraints, and,
seamlessly relates canonical probability distributions obtained
using the normal averages, C-T, $q$-averages, and OLM constraints.

The normal averages constraints [1] were initially abandoned because
of difficulties encountered in obtaining an acceptable form for the
partition function, and, the nonextensive statistics community have largely utilized $q$-averages constraints instead of the C-T constraints, since $ \left\langle 1 \right\rangle _q \ne
1 $. Recent studies by Abe [19,20] suggest that unlike $q$-averages,
normal averages are physical and consistent with both the
generalized H-theorem and the generalized \textit{Stosszahlansatz} (molecular
chaos hypothesis).

Despite this physics-related deficiency prominently cited prior
works on GVPA's [8-10] utilized the C-T constraints as a consequence
of mathematical necessity. For C-T constraints, the generalized
internal energy: $ U _q = <E_n>_q= \sum\limits_n {p_n^q } E_n \ $.
The canonical distribution that maximizes the Tsallis entropy is
[8-10,15]: $p_n  = p\left( {E_n } \right) = \frac{{\left[ {1 -
\left( {1 - q} \right)\beta E_n } \right]^{\frac{1}{{1 - q}}}
}}{{\tilde Z\left( \beta  \right)}}$. The partition function is: $
\tilde Z\left( \beta  \right) = \sum\limits_n {\left[ {1 - \left( {1
- q} \right)\beta E_n } \right]^{\frac{1}{{1 - q}}} }$, where, $
\beta $ is the Lagrange multiplier associated with the generalized
internal energy $ U_q  $.

From (1), the GFE is: $ F_q  =  - \frac{1}{\beta }\frac{{\tilde
Z\left( \beta  \right)^{1 - q}  - 1}}{{1 - q}} $. Thus, the
first-order Newtonian derivative of the GFE is of the form: $
\frac{{dF_q }}{{d\lambda }} =  \tilde Z\left( \beta \right)^{ - q}
\sum\limits_n {\left[ {1 - \left( {1 - q} \right)\beta E_n }
\right]^{\frac{q}{{1 - q}}} \frac{{dE_n }}{{d\lambda }}} $.  The
first-order Newtonian derivative of the GFE naturally yields:
$\frac{{dF_q }}{{d\lambda }}= \sum\limits_n {p_n^q \frac{{dE_n
}}{{d\lambda
 }}} =
\left\langle {\frac{{dE_n }}{{d\lambda }}} \right\rangle _q  $. The
mathematical expression for the \textit{q-deformed }exponential: $
\exp _q x = \left[ {1 + \left( {1 - q} \right)x}
\right]^{\frac{1}{{1 - q}}} $ [21] which, satisfies the differential
equation: $ \frac{{dy}}{{dx}} = y^q $, mandates that Newtonian
derivatives of the GFE (1) results in expectations defined in the
C-T form. These strictures severely constrain the
scope and generality of the GVPA analyses described in [8-10].

The primary leitmotif of this paper is to derive a candidate
\textit{q-deformed} GVPA employing the \textit{q-deformed}
derivatives defined by [21]: $ D_{\left( q \right)}^x y = \left[ {1
+ \left( {1 - q} \right)x} \right]\frac{{dy}}{{dx}} $, where the
\textit{q-deformed }exponential satisfies:$ D_{\left( q \right)}^x y
= y $.  One of the significant consequences of such a generalization
is the ability to define expectations defined in terms of normal
averages, when taking derivatives of the GFE.

It has been suggested [10] that the results of Ref. [18] could be employed to transform
GVPA's derived using C-T constraints to equivalent forms described
by either normal averages or $ q $-averages.  While such a
suggestion may be true in principle for simple cases, its practical
tractability is questionable when applied to studies in
communication theory and machine learning, where the variational
extremization of the maximum Tsallis entropy problem is done with
respect to conditional (transition) probabilities.  Furthermore,
this prescription does not result in a true generalization of the
variational procedure employed in GVPA's, within the framework of
\textit{q-deformed} calculus [21].

In accordance with [18], a generic form of the canonical probability
distribution that maximizes the Tsallis entropy is
\begin{equation}
\begin{array}{l}
 p_n  = \frac{{\left[ {1 - \left( {1 - q^ *  } \right)\beta ^ *  E_n } \right]^{\frac{1}{{1 - q^ *  }}} }}{{Z\left( {\beta ^ *  } \right)}};\sum\limits_n {p_n  = 1,}  \\
 \tilde Z\left( {\beta ^ *  } \right) = \sum\limits_n {\left[ {1 - \left( {1 - q^ *  } \right)\beta ^ *  E_n } \right]^{\frac{1}{{1 - q^ *  }}} .}  \\
 \end{array}
\end{equation}
For normal averages constraints
\begin{equation}
\begin{array}{l}
 q^ *   = 2 - q \\
 \beta ^ *=
\frac{{{\beta  \mathord{\left/
 {\vphantom {\beta  q}} \right.
 \kern-\nulldelimiterspace} q}}}{{\aleph _q  + \frac{{\left( {q - 1} \right)}}{q}\beta U_q }} = \frac{{\tilde \beta }}{{\aleph _q  + \left( {q - 1} \right)\tilde \beta U_q }};where:\tilde \beta  = {\raise0.7ex\hbox{$\beta $} \!\mathord{\left/
 {\vphantom {\beta  q}}\right.\kern-\nulldelimiterspace}
\!\lower0.7ex\hbox{$q$}}, \\
 \aleph _q  = \sum\limits_n {p_n^q ,and,} U_q  = \sum\limits_n {p_n E_n .}  \\
 \end{array}
\end{equation}
For the C-T constraints
\begin{equation}
\begin{array}{l}
 q^ *   = q \\
 \beta ^ *   = \beta , \\
 U_q  = \sum\limits_n {p_n^q } E_n . \\
 \end{array}
\end{equation}
For normal averages constraints, (6) and (7) employ the
\textit{additive duality} of generalized statistics (see [1] and the references therein).
Specifically, $ q\rightarrow 2-q=q^* $.\footnote{Note that
$q\rightarrow q^*$ denotes re-parameterization from $ q $ to $ q^*
.$}  Here, (6) is subjected to the Tsallis cut-off condition
[1]: $ \left[ {1 - \left( {1 - q^ *  } \right)\beta ^ *  E_n }\right] \le 0 $.  The rationale for denoting the generalized internal energy as: $U_q$, is to highlight its nonextensive nature irrespective of the form of expectation involved in its definition. Note that even for expectations defined by normal averages, $U_q$ has an implicit $q$-dependency facilitated by the canonical probability (6).

While replacing the Shannon entropy with the Tsallis entropy
yielding GVPA's [8-10] represents an initial level of
generalization, the \textit{q-deformed} GVPA model described in this
paper achieves a further level of generalization by replacing the
Newtonian derivative with the \textit{q-deformed} derivative in the
variational procedure. Specifically, given a function: $ F\left(
\tau \right) = \sum\limits_n {F\left( {\tau _n } \right)} $, the
chain rule yields: $ \frac{{dF \left( \tau \right)}}{{d\lambda }} =
\frac{{dF \left( \tau \right)}}{{d\tau }}\frac{{d\tau }}{{d\lambda
}} \ $. Replacing the Newtonian derivative: $ \frac{{dF \left( \tau
\right)}}{{d\tau }} $ by the \textit{q-deformed} derivative defined
by [21]:  $ D^\tau_{\left( q \right)}F(\tau) = \left[ {1 + \left( {1
- q} \right)\tau } \right]\frac{{dF\left( \tau \right)}}{{d\tau }} $
(see Section 2)and defining: $ D_{\left( q \right)}^\tau F \left(
\tau \right)\frac{{d\tau }}{{d\lambda }} =\delta _{\left( q
\right),\tau} F \left( \tau \right) $, facilitates the
transformation: $ \frac{{dF \left( \tau \right)}}{{d\lambda }} \to
\delta _{\left( q \right),\tau} F \left( \tau \right) $.  \emph{Note
that the \textit{q-deformed} derivative $ D^\tau_{(q)} $ operates
only on the term $ F(\tau) $, which in this paper is a thermodynamic
function expressed as a q-deformed exponential.  Within the scope of
this paper, $ F(\tau) $ comprises the GFE}. Thus the increasing
order to derivatives acquire the form
\begin{equation}
\begin{array}{l}
\delta _{\left( q \right),\tau} F(\tau) = D^\tau_{\left( q \right)}  F(\tau)\frac{{d\tau }}{{d\lambda }}, \\
 \delta _{\left( q \right),\tau}^2 F\left( \tau  \right) = D_{\left( q \right)}^\tau \frac{{d\tau }}{{d\lambda }}\delta _{\left( q \right),\tau} F\left( \tau  \right), \\
...... \\
\delta _{\left( q \right),\tau}^{k + 1} F\left( \tau  \right) =
D_{\left( q \right)}^\tau  \frac{{d\tau }}{{d\lambda }}\delta
_{\left( q
\right),\tau}^k F\left( \tau  \right);k = 1,.... \\
\end{array}
\end{equation}
This results in \textit{q-deformed} GVPA models that simultaneously
exhibit generalization within the context of the generalized
statistics framework employed, and, in their underlying mathematical
structure. The commutation relations: $ D^\tau_{\left( q \right)}
F\left( \tau \right) = \sum\limits_n {D^{\tau_n}_{\left( q \right)}
F\left( {\tau _n } \right)} $, and, $ \delta _{\left( q \right),\tau
} F\left( \tau  \right) = D_{\left( q \right)}^\tau  F\left( \tau
\right)\frac{{d\tau }}{{d\lambda }} = \sum\limits_n {D_{\left( q
\right)}^{\tau _n } F\left( {\tau _n } \right)\frac{{d\tau _n
}}{{d\lambda }}}  = \sum\limits_n {\delta _{\left( q \right),\tau _n
} F\left( {\tau _n } \right)}  $ are established in Theorem 1. These
relations are critical to the correctness and admissability of the
\textit{q-deformed} GVPA model, and are derived in Section 2 of this
paper.

Section 3 derives a \textit{q-deformed} GVPA model employing the
\textit{q-deformed} derivative in the variational
procedure. This is accomplished with the aid of the Hellmann-Feynman
theorem [12]. Section 3 also analyzes the commutation between the
Newtonian derivative $ \frac{d}{{d\lambda }}$ and the summation sign, and the contribution of the cut-off in the variational and perturbation methods for higher-order perturbation terms.  \textit{It is noteworthy to
mention that expectations in terms of normal averages are achieved
by relating the "inverse thermodynamic temperature" $\beta$ (the
energy Lagrange multiplier) to the "physical inverse temperature":
$\beta^*$ [13] via (7), as is demonstrated in Section 3}.  This is accomplished through the relation in (7): $ \tilde\beta=\beta^*\tilde Z(\beta^*)^{q^*-1}$, where: $\tilde\beta=\beta/q$ is the scaled "inverse thermodynamic temperature".

Section 4 formulates the generalized Bogoliubov inequality [5,7-10]
truncated at first-order terms for the case of the classical
harmonic oscillator. Numerical examples demonstrating the results
for the \textit{q-deformed} GVPA are presented in Section 5.
\textit{The ability of the q-deformed GVPA model presented in this
paper to demonstrate both sub-extensivity (sub-additivity) and super-extensivity (super-additivity) within
the context of the generalized Bogoliubov inequality truncated at
first-order terms is demonstrated}.  This feature is not possessed
by existing GVPA models [8, 10]. Section 6 concludes this paper.

\section{Theoretical preliminaries}
\subsection{Tsallis entropy}

The \textit{q-deformed} logarithm and exponential are defined as
[21]
\begin{equation}
\begin{array}{l}
\ln _q \left( x \right) = \frac{{x^{1 - q} - 1}}{{1 - q}}, \\
and, \\
\exp _q \left( x \right) = \left\{ \begin{array}{l}
 \left[ {1 + \left( {1 - q} \right)x} \right]^{\frac{1}{{1 - q}}} ;1 + \left( {1 - q} \right)x \ge 0 \\
 0;othewise, \\
 \end{array} \right.
\end{array}
\end{equation}
respectively. By definition, the un-normalized Tsallis entropy, is
defined in terms of discrete variables as [1, 2]
\begin{equation}
S_q  =  - \frac{{1 - \sum\limits_n {p_n^q } }}{{1 - q}} =  -
\sum\limits_n {p_n^q } \ln _q p_n ;\sum\limits_n {p_n }  = 1.
\end{equation}
The constant $ q $ is referred to as the nonadditivity parameter.

\subsection{Results from q-algebra and q-calculus}

The \textit{q-deformed} addition $ \oplus_q $ and the
\textit{q-deformed} subtraction $ \ominus_q $ are defined as
\begin{equation}
\begin{array}{l}
 x \oplus_q y = x + y + \left( {1 - q} \right)xy, \\
  \ominus_q y = \frac{{ - y}}{{1 + \left( {1 - q} \right)y}};1+(1-q)y > 0 \Rightarrow x \ominus_q y = \frac{{ x- y}}{{1 + \left( {1 - q} \right)y}} \\
 \end{array}
\end{equation}

The \textit{q-deformed} derivative, is defined as
\begin{equation}
D^\tau_{\left( q \right)} F\left( \tau  \right) = \mathop {\lim
}\limits_{\nu  \to \tau } \frac{{F\left( \tau  \right) - F\left( \nu
\right)}}{{\tau  \ominus_q \nu }} = \left[ {1 + \left( {1 - q}
\right)\tau } \right]\frac{{dF\left( \tau  \right)}}{{d\tau }}
\end{equation}

\subsection{Deformed calculus framework for \textit{q-deformed} GVPA}

Consider a function: $ F\left( {g\left( E \right)} \right) =
\sum\limits_n {\left[ {1 - \left( {1 - q} \right)\beta g\left( {E_n
} \right)} \right]^{\frac{1}{{1 - q}}} }  = \sum\limits_n {F\left(
{g\left( {E_n } \right)} \right)} $. On comparison with (6), $
F(\tau) $ may be construed as a generic form of the canonical
partition function $\tilde Z(\beta) $.  Substituting $
 - \beta g(E) = \tau $, yields $
F\left( \tau  \right) = \sum\limits_n {\left[ {1 + \left( {1 - q}
\right)\tau _n } \right]^{\frac{1}{{1 - q}}} }  = \sum\limits_n
{F\left( {\tau _n } \right)} $.

\textbf{Theorem 1}:  Given a function $ F\left( \tau  \right) =
\sum\limits_n {\left[ {1 + \left( {1 - q} \right)\tau _n }
\right]^{\frac{1}{{1 - q}}} }  = \sum\limits_n {F\left( {\tau _n }
\right)} $, where $ \tau = \left\{ {\tau _1 ,.......,\tau _N }
\right\} $ are $ N $ separate instances of $ \tau_n ; n=1,...,N$.
The following relation involving action of the \textit{q-deformed}
derivative
\begin{equation}
D^\tau_{\left( q \right)} F\left( \tau  \right) = \sum\limits_n {D^{\tau_n}_{\left( q \right)} \left[ {1 + \left( {1 - q} \right)\tau _n } \right]^{\frac{1}{{1 - q}}} }  = \sum\limits_n {D^{\tau_n}_{\left( q \right)} F\left( {\tau _n } \right)},  \\
\end{equation}
and, \\
\begin{equation}
\begin{array}{l}
D_{\left( q \right)}^\tau  F\left( \tau  \right)\frac{{d\tau
}}{{d\lambda }} = \sum\limits_n {D_{\left( q \right)}^{\tau _n }
\left[ {1 + \left( {1 - q} \right)\tau _n } \right]^{\frac{1}{{1 -
q}}} \frac{{d\tau _n }}{{d\lambda }}}  = \sum\limits_n {D_{\left( q
\right)}^{\tau _n } F\left( {\tau _n } \right)\frac{{d\tau _n
}}{{d\lambda }}}.\\
 \end{array}
\end{equation}
hold true.

\textbf{Proof}:  Employing (13) yields
\begin{equation}
\begin{array}{l}
 D^\tau_{\left( q \right)} F\left( \tau  \right) = \left[ {1 + \left( {1 - q} \right)\tau } \right]\frac{{dF\left( \tau  \right)}}{{d\tau }} \\
  = \left[ {1 + \left( {1 - q} \right)\tau _1 } \right]\frac{{dF\left( {\tau _1 } \right)}}{{d\tau _1 }} + ... + \left[ {1 + \left( {1 - q} \right)\tau _N } \right]\frac{{dF\left( {\tau _N } \right)}}{{d\tau _N }} \\
  = D_{\left( q \right)}^{\tau _1 } F\left( {\tau _1 } \right) + ... + D_{\left( q \right)}^{\tau _N } F\left( {\tau _N } \right) = \sum\limits_n {D_{\left( q \right)}^{\tau _n } F\left( {\tau _n } \right)}.  \\
 \end{array}
\end{equation}

Similarly,
\begin{equation}
\begin{array}{l}
 D_{\left( q \right)}^\tau  F\left( \tau  \right)\frac{{d\tau }}{{d\lambda }} \\
  = \left[ {1 + \left( {1 - q} \right)\tau _1 } \right]\frac{{dF\left( {\tau _1 } \right)}}{{d\tau _1 }}\frac{{d\tau _1 }}{{d\lambda }} + ... \\
  + \left[ {1 + \left( {1 - q} \right)\tau _N } \right]\frac{{dF\left( {\tau _N } \right)}}{{d\tau _N }}\frac{{d\tau _N }}{{d\lambda }} \\
  = \frac{d}{{d\lambda }}\sum\limits_n {D_{\left( q \right)}^{\tau _n } F\left( {\tau _n } \right)\tau _n }  = \sum\limits_n {D_{\left( q \right)}^{\tau _n } F\left( {\tau _n } \right)\frac{{d\tau _n }}{{d\lambda }}} . \\
 \end{array}
\end{equation}

Note that (17) assumes that the Newtonian derivative $
\frac{d}{{d\lambda }} $ commutes with the summation sign [8].  The
results of Theorem 1 form the basis for the \textit{q-deformed} GVPA
model presented in this paper, which extends earlier GVPA studies
[8-10]. Note that (17) may be extended to obtain expressions for
higher order derivatives with the aid of (9), which is demonstrated
in Section 3. Setting: $ F\left( \tau \right) = \left[ {1 + \left(
{1 - q} \right)\tau } \right]^{\frac{1}{{1 - q}}} = \sum\limits_n
{\left[ {1 + \left( {1 - q} \right)\tau _n } \right]^{\frac{1}{{1 -
q}}} } $, $ D^\tau_{(q)}F(\tau) $ yields with the aid of (16)
\begin{equation}
\begin{array}{l}
 D_{\left( q \right)}^\tau  F\left( \tau  \right) = \left[ {1 + \left( {1 - q} \right)\tau } \right]\frac{{d\left[ {1 + \left( {1 - q} \right)\tau } \right]^{\frac{1}{{1 - q}}} }}{{d\tau }} \\
  = \sum\limits_n {\left[ {1 + \left( {1 - q} \right)\tau _n } \right]\left[ {1 + \left( {1 - q} \right)\tau _n } \right]^{\frac{q}{{1 - q}}} } \\
   = \sum\limits_n {\left[ {1 + \left( {1 - q} \right)\tau _n } \right]^{\frac{1}{{1 - q}}} }  = \sum\limits_n {F\left( {\tau _n } \right)}.  \\
 \end{array}
\end{equation}
Analogously
\begin{equation}
\frac{{dF\left( \tau  \right)}}{{d\tau }} = \sum\limits_n {\left[ {1
+ \left( {1 - q} \right)\tau _n } \right]^{\frac{q}{{1 - q}}}}.\\
\end{equation}

Setting $\tau=-\beta E\Rightarrow \frac{{d\tau }}{{dE}} =  - \beta
$, $ \delta _{(q),\tau} F\left( \tau  \right) = D_{\left( q
\right)}^\tau F\left( \tau  \right)\frac{{d\tau }}{{d\lambda }} =
D_{\left( q \right)}^\tau  F\left( \tau  \right)\frac{{d\tau
}}{{dE}}\frac{{dE}}{{d\lambda }} =  - \beta D_{\left( q
\right)}^\tau  F\left( \tau  \right)\frac{{dE}}{{d\lambda }} $.
Thus, (17) and (18) yield
\begin{equation}
\begin{array}{l}
 \left. {\delta _{\left( q \right),\tau } F\left( \tau  \right)} \right|_{\lambda  = 0}  = \left. {\left[ {1 - \left( {1 - q} \right)\beta E} \right]\frac{{dF\left( E \right)}}{{dE}}\frac{{dE}}{{d\lambda }}} \right|_{\lambda  = 0} \\
 = -\beta\left. {\sum\limits_n {\left[ {1 - \left( {1 - q} \right)\beta E_n } \right]} ^{\frac{1}{{1 - q}}} \frac{{dE_n }}{{d\lambda }}} \right|_{\lambda  = 0}.  \\
 \end{array}
\end{equation}
Analogously
\begin{equation}
\left. {\frac{{dF\left( \tau  \right)}}{{d\lambda }}}
\right|_{\lambda  = 0}  = -\beta\left. {\sum\limits_n {\left[ {1 -
\left( {1 - q} \right)\beta E_n } \right]} ^{\frac{q}{{1 - q}}}
\frac{{dE_n }}{{d\lambda }}} \right|_{\lambda  = 0}.
\end{equation}

Comparison between (20) and (21) provides initial evidence that
optimality conditions obtained using the \textit{q-deformed}
derivative qualitatively differ from those obtained using the
Newtonian derivative.

\section{q-Deformed generalized variational perturbation approximation}

\subsection{Overview of the Rayleigh-Schr\"{o}dinger perturbation method}
Consider the eigenvalue equation
\begin{equation}
H\left| {n} \right\rangle  = E_n \left| {n }
\right\rangle, \\
\end{equation}
where $ H$ is the Hamiltonian, $ E_n $ is the energy eigenvalue, $ n
$ is the quantum number, and, $ \left| n \right\rangle  $ is the
eigenfunction (eigenvector).  Note that the non-degenerate case is
considered herein.  The eigenfunctions and the eigenvalues are
written as [22]
\begin{equation}
\begin{array}{l}
\left| n \right\rangle  = \left| n \right\rangle ^{\left( 0 \right)}
+ \lambda \left| n \right\rangle ^{\left( 1 \right)}  + \lambda ^2
\left| n \right\rangle ^{\left( 2 \right)}  + ..., \\
 E_n  = E_n^{(0)}  + \lambda E_n^{(1)}  + \lambda ^2 E_n^{(2)}  + ..., \\
 \end{array}
\end{equation}
where $ \lambda \in [0,1] $.  In (23), $ E_n^{(k)} $ and $ \left| n
\right\rangle ^{\left( k \right)} $ are the $ k^{th} $ corrections
to the energy eigenvalues and eigenfunctions, respectively.  Note
that the non-degenerate theory is valid only when: $ \left|
{{}^{\left( 0 \right)}\left\langle m \right|H_1 \left| n
\right\rangle ^{\left( 0 \right)} } \right|  < < \left| {E_n^{(0)} -
E_m^{(0)} } \right| $. Substituting (23) into the time-independent
Schr\"{o}dinger equation, and comparing the coefficients of the
powers of $\lambda $ on either side, the first two eigenfunction
corrections are
\begin{equation}
\begin{array}{l}
 \left| n \right\rangle ^{\left( 1 \right)}  = \sum\limits_{m \ne n} {\frac{{{}^{\left( 0 \right)}\left\langle m \right|H_1 \left| n \right\rangle ^{\left( 0 \right)} }}{{E_n^{(0)}  - E_m^{(0)} }}} \left| m \right\rangle ^{\left( 0 \right)}  = \sum\limits_{m \ne n} {\frac{{H_{1_{mn} } }}{{E_n^{(0)}  - E_m^{(0)} }}} \left| m \right\rangle ^{\left( 0 \right)}  \\
 \left| n \right\rangle ^{\left( 2 \right)}  = \sum\limits_{m \ne n} {\left[ {\sum\limits_{l \ne n} {\frac{{H_{1_{ml} } H_{1_{ln} } }}{{\left( {E_n^{(0)}  - E_m^{(0)} } \right)\left( {E_n^{(0)}  - E_l^{(0)} } \right)}} - \frac{{H_{1_{mn} } H_{1_{nn} } }}{{\left( {E_n^{(0)}  - E_m^{(0)} } \right)^2 }}} } \right]} \left| m \right\rangle ^{\left( 0 \right)}  \\
  - \frac{1}{2}\sum\limits_{m \ne n} {\frac{{\left( {H_{1_{mn} } } \right)^2 }}{{\left( {E_n^{(0)}  - E_m^{(0)} } \right)^2 }}} \left| n \right\rangle ^{\left( 0 \right)}.  \\
 \end{array}
\end{equation}

\subsection{Generalized free energy correction terms}
Setting $ \tau _n  =  - \beta ^ *  E_n $, (6) acquires the form
\begin{equation}
\begin{array}{l}
 p_n  = \frac{{\left[ {1 + \left( {1 - q^*} \right)\tau _n } \right]^{\frac{1}{{1 - q^ *  }}} }}{{\tilde Z\left( \tau  \right)}}, \\
 \tilde Z\left( \tau  \right) = \sum\limits_n {\left[ {1 + \left( {1 - q^*} \right)\tau _n } \right]^{\frac{1}{{1 - q^ *  }}} } . \\
 \end{array}
\end{equation}
For expectations defined by normal averages, the GFE (1) is defined by: $F_q = U_q  - \frac{1}{{\tilde \beta }}S_q =  - \frac{1}{{\tilde \beta }}\ln _q \tilde Z\left( {\tilde\beta } \right) $, where: $\tilde\beta=\beta/q$.  Note that the scaled energy Lagrange multiplier $ \tilde\beta $ is introduced in order to achieve consistency between the two forms of the GFE. The \textit{$q^*$-deformed}  GFE is defined as
\begin{equation}
\Im _{q^ *  }  = - \frac{1}{\tilde\beta }\frac{{\tilde Z\left( \tau
\right)^{q^ *-1  }  - 1}}{{q^ *-1  }}.
\end{equation}
The derivation of (26) is provided in Appendix A of this paper.
Defining
\begin{equation}
\delta _{\left( q^* \right),\tau } \tilde Z\left( \tau \right) =
D_{\left( q^* \right)}^\tau  \tilde Z\left( \tau \right)\frac{{d\tau
}}{{d\lambda }},
\end{equation}
a perturbation expansion of the $q^*-deformed $ GFE, analogous to
(3), is

\begin{equation}
\begin{array}{l}
 \Im _{q^ *  } \left( \lambda  \right) = \Im _{q^ *  }^{(0)}  + \left. {\lambda \Im _{q^ *  }^{\left( 1 \right)} } \right|_{\lambda  = 0}  + \frac{{\lambda ^2 }}{{2!}}\left. {\Im _{q^ *  }^{\left( 2 \right)} } \right|_{\lambda  = 0}  + ... \\
\Im _{q^ *  }^{\left( 1 \right)}  = \delta _{\left( {q^ * }
\right),\tau } \Im _{q^ *  }, \\
\Im _{q^ *  }^{\left( 2 \right)}  = \delta _{\left( {q^ *  }
\right),\tau } \Im _{q^ *  }^{\left( 1 \right)},\\
...........,\\
\Im _{q^ *  }^{\left( {k + 1} \right)}  = \delta _{\left( {q^ *  }
\right),\tau } \Im _{q^ *  }^{\left( k \right)}
\end{array}
 \end{equation}
\subsection{Derivation of terms in the GFE perturbation expansion}
 From (23) and (26), the zeroth-order term is
\begin{equation}
\begin{array}{l}
 \Im _{q^ *  }^{\left( 0 \right)}  =  - \frac{1}{{\tilde\beta   }}\frac{{\tilde Z_{\left( {q^ *  } \right),0} \left( {\beta ^ *  } \right)^{q^ *-1  }  - 1}}{{ q^ *-1  }}, \\
 where \\
 \tilde Z_{\left( {q^ *  } \right),0} \left( {\beta ^ *  } \right) = \sum\limits_n {\left[ {1 - \left( {1 - q^ *  } \right)\beta ^ *  E_n^{\left( 0 \right)} } \right]^{\frac{1}{{1 - q^ *  }}} } . \\
 \end{array}
\end{equation}

Given the canonical probability distribution and the canonical
partition function in (25), (28) yields
\begin{equation}
\begin{array}{l}
 \Im _{q^ *  }^{\left( 1 \right)}  = \delta _{\left( {q^ *  } \right),\tau } \Im _{q^ *  }  = D_{q^ *  }^\tau  \Im _{q^ *  } \left( {\tilde Z\left( \tau  \right)} \right)\frac{{d\tau }}{{d\lambda }} \\
  = \left[ {1 + \left( {1 - q^ *  } \right)\tau } \right]\frac{{d\Im _{q^ *  } \left( {\tilde Z\left( \tau  \right)} \right)}}{{d\tau }}\frac{{d\tau }}{{d\lambda }} \\
  = \left[ {1 + \left( {1 - q^ *  } \right)\tau } \right]\frac{{d\Im _{q^ *  } \left( {\tilde Z\left( \tau  \right)} \right)}}{{d\tilde Z\left( \tau  \right)}}\frac{{d\tilde Z\left( \tau  \right)}}{{d\tau }}\frac{{d\tau }}{{d\lambda }}. \\
\end{array}
\end{equation}

Substituting (25) and (26) into (30), and employing (15), yields
\begin{equation}
\begin{array}{l}
\Im _{q^ *  }^{\left( 1 \right)}=  - \frac{1}{\tilde\beta }\tilde Z\left( \tau  \right)^{q^ *   - 2} \sum\limits_n {\left[ {1 + \left( {1 - q^ *  } \right)\tau _n } \right]\frac{d}{{d\tau _n }}\left[ {1 + \left( {1 - q^ *  } \right)\tau _n } \right]^{\frac{1}{{1 - q^ *  }}} } \frac{{d\tau _n }}{{d\lambda }} \\
  =  - \frac{1}{\tilde\beta }\tilde Z\left( \tau  \right)^{q^ *   - 2} \sum\limits_n {\left[ {1 + \left( {1 - q^ *  } \right)\tau _n } \right]^{\frac{1}{{1 - q^ *  }}} } \frac{{d\tau _n }}{{d\lambda }}. \\
\end{array}
\end{equation}

Setting: $ \tau_n=-\beta^*E_n$, employing the chain rule: $
\frac{{d\tau _n }}{{d\lambda }} = \frac{{d\tau _n }}{{dE_n
}}\frac{{dE_n }}{{d\lambda }} $, and substituting: $ \tilde \beta=
\beta^* \tilde Z(\beta^*)^{q^*-1} $ from (7) into (31) yields
\begin{equation}
\Im _{q^ *  }^{\left( 1 \right)}= \frac{1}{{\tilde Z\left( {\beta ^ *  } \right)}}\sum\limits_n {\left[ {1 - \left( {1 - q^ *  } \right)\beta ^ *  E_n } \right]^{\frac{1}{{1 - q^ *  }}} } \frac{{dE_n }}{{d\lambda }}. \\
\end{equation}

Employing the Hellmann-Feynman theorem (4), the first-order
perturbation term in (28) is defined as
\begin{equation}
\begin{array}{l}
 \left. {\Im _{q^ *  }^{\left( 1 \right)} } \right|_{\lambda  = 0}  = \left. {\frac{1}{{\tilde Z\left( {\beta ^ *  } \right)}}\sum\limits_n {\left[ {1 - \left( {1 - q^ *  } \right)\beta ^ *  E_n } \right]} ^{\frac{1}{{1 - q^ *  }}} \frac{{dE_n }}{{d\lambda }}} \right|_{\lambda  = 0}  \\
  = \left. {\frac{1}{{\tilde Z\left( {\beta ^ *  } \right)}}\sum\limits_n {\left[ {1 - \left( {1 - q^ *  } \right)\beta ^ *  E_n } \right]} ^{\frac{1}{{1 - q^ *  }}} H_{1_{nn} } } \right|_{\lambda  = 0}  \\
  = \frac{1}{{\tilde Z_{\left( {q^ *  } \right),0} \left( {\beta ^ *  } \right)}}\sum\limits_n {\left[ {1 - \left( {1 - q^ *  } \right)\beta ^ *  E_n^{\left( 0 \right)} } \right]} ^{\frac{1}{{1 - q^ *  }}} H_{1_{nn} }  \\
  = \sum\limits_n {p_n \left( {E_n^{\left( 0 \right)} } \right)H_{1_{nn} } }  = \left\langle {H_1 } \right\rangle _{q^ *  }^{\left( 0 \right)} , \\
 \end{array}
\end{equation}
\textit{where, $\left\langle {H_1 } \right\rangle _{q^ *  }^{\left( 0
\right)}$ denotes the expectation of $H_1$ in normal averages form,
defined with respect to the probability $p(E_n^{(0)})$ parameterized
by $q^*$}.

Note that $ \Im _{q^ *  }^{\left( 1 \right)} $in (32) is obtained by
operating on the \textit{$q^*$-deformed} GFE in (26) with $ \delta
_{\left( {q^ *  } \right),\tau _n } $, where $ \tau_n=-\beta^*E_n $.
Thus, the "inverse thermodynamic temperature" $\beta$ (the energy
Lagrange multiplier) is related to the "physical inverse
temperature" $\beta^* $ by the relation:
$\beta/q=\tilde\beta=\beta^*\tilde Z(\beta^*)^{q^*-1}$, employing
the results of Theorem 1. In (33), $ \left. {\Im _{q^
*  }^{\left( 1 \right)} } \right|_{\lambda  = 0} $ is tantamount to specifying the $E_n=E_n^{(0)}$ as defined in (4) and (23).  Thus, $\tilde\beta \ne \beta^*\tilde Z_{(q^*),0}(\beta^*)^{q^*-1}$, where $\tilde Z_{(q^*),0}(\beta^*)$ is defined in (29).

Setting $ \tau_n=-\beta^*E_n $, (32) is re-expressed as
\begin{equation}
\Im _{q^ *  }^{\left( 1 \right)} =- \frac{1}{{\beta ^ *  }}\left\{ {\tilde Z\left( \tau  \right)^{ - 1} \sum\limits_n {\left[ {1 + \left( {1 - q^ *  } \right)\tau _n } \right]^{\frac{1}{{1 - q^ *  }}} } \frac{{d\tau _n }}{{d\lambda }}} \right\}. \\
 \end{equation}

Operating on (34) with $ \delta_{(q^*),\tau} $ defined in (28),
yields

\begin{equation}
\begin{array}{l}
 \Im _{q^ *  }^{\left( 2 \right)}  = \delta _{\left( {q^ *  } \right),\tau } \Im _{q^ *  }^{\left( 1 \right)}  = D_{(q^ *)  }^\tau  \Im _{q^ *  }^{\left( 1 \right)} \frac{{d\tau }}{{d\lambda }} \\
  =  - \frac{1}{{\beta ^ *  }}D_{(q^ *)  }^\tau  \left\{ {\tilde Z\left( \tau  \right)^{ - 1} \sum\limits_n {\left[ {1 + \left( {1 - q^ *  } \right)\tau _n } \right]^{\frac{1}{{1 - q^ *  }}} } \frac{{d\tau _n }}{{d\lambda }}} \right\}\frac{{d\tau }}{{d\lambda }}. \\
\end{array}
 \end{equation}

Employing the Leibnitz rule for \emph{deformed} derivatives [21]
\begin{equation}
D_{\left( {q^ *  } \right)}^\tau  \left[ {A\left( \tau
\right)B\left( \tau  \right)} \right] = B\left( \tau
\right)D_{\left( {q^ *  } \right)}^\tau  A\left( \tau  \right) +
A\left( \tau \right)D_{\left( {q^ *  } \right)}^\tau  B\left( \tau
\right),
 \end{equation}

and defining: $ A\left( \tau  \right) =  \sum\limits_n {\left[ {1 +
\left( {1 - q^ * } \right)\tau _n } \right]^{\frac{1}{{1 - q^ * }}}
\frac{{d\tau _n }}{{d\lambda }}} $ and $ B\left( \tau \right)=\tilde
Z\left( \tau \right)^{ - 1} $, yields

\begin{equation}
\begin{array}{l}
 \Im _{q^ *  }^{\left( 2 \right)}  = \underbrace { - \frac{1}{{\beta ^ *  }}\left\{ {\tilde Z\left( \tau  \right)^{ - 1} D_{(q^ *)  }^\tau  \sum\limits_n {\left[ {1 + \left( {1 - q^ *  } \right)\tau _n } \right]^{\frac{1}{{1 - q^ *  }}} } \frac{{d\tau _n }}{{d\lambda }}} \right\}\frac{{d\tau }}{{d\lambda }}}_{Term - 1} \\
 \underbrace { - \frac{1}{{\beta ^ *  }}\left\{ {\sum\limits_n {\left[ {1 + \left( {1 - q^ *  } \right)\tau _n } \right]^{\frac{1}{{1 - q^ *  }}} } \frac{{d\tau _n }}{{d\lambda }}D_{(q^ *)  }^\tau  \tilde Z\left( \tau  \right)^{ - 1} } \right\}\frac{{d\tau }}{{d\lambda }}}_{Term - 2}. \\
\end{array}
 \end{equation}

Evaluating (37) term-wise with the aid of (15), yields
\begin{equation}
\begin{array}{l}
 Term - 1 =  - \frac{1}{{\beta ^ *  }}\left\{ {\tilde Z\left( \tau  \right)^{ - 1} D_{\left( {q^ *  } \right)}^\tau  \sum\limits_n {\left[ {1 + \left( {1 - q^ *  } \right)\tau _n } \right]^{\frac{1}{{1 - q^ *  }}} } \frac{{d\tau _n }}{{d\lambda }}} \right\}\frac{{d\tau }}{{d\lambda }} \\
  =  - \frac{1}{{\beta ^ *  }}\left\{ {\tilde Z\left( \tau  \right)^{ - 1} \sum\limits_n {\left[ {1 + \left( {1 - q^ *  } \right)\tau _n } \right]^{\frac{1}{{1 - q^ *  }}} } \left( {\frac{{d\tau _n }}{{d\lambda }}} \right)^2 } \right. \\
  + \left. {\tilde Z\left( \tau  \right)^{ - 1} \sum\limits_n {\left[ {1 + \left( {1 - q^ *  } \right)\tau _n } \right]^{\frac{{2 - q^ *  }}{{1 - q^ *  }}} \frac{d}{{d\tau _n }}\left( {\frac{{d\tau _n }}{{d\lambda }}} \right)\left( {\frac{{d\tau _n }}{{d\lambda }}} \right)} } \right\} \\
  =  - \frac{1}{{\beta ^ *  }}\left\{ {\tilde Z\left( \tau  \right)^{ - 1} \sum\limits_n {\left[ {1 + \left( {1 - q^ *  } \right)\tau _n } \right]^{\frac{1}{{1 - q^ *  }}} } \left( {\frac{{d\tau _n }}{{d\lambda }}} \right)^2 } \right. \\
  + \left. {\tilde Z\left( \tau  \right)^{ - 1} \sum\limits_n {\left[ {1 + \left( {1 - q^ *  } \right)\tau _n } \right]^{\frac{{2 - q^ *  }}{{1 - q^ *  }}} \frac{{d\lambda }}{{d\tau _n }}\frac{d}{{d\lambda }}\left( {\frac{{d\tau _n }}{{d\lambda }}} \right)\left( {\frac{{d\tau _n }}{{d\lambda }}} \right)} } \right\} \\
  =  - \frac{1}{{\beta ^ *  }}\left\{ {\tilde Z\left( \tau  \right)^{ - 1} \sum\limits_n {\left[ {1 + \left( {1 - q^ *  } \right)\tau _n } \right]^{\frac{1}{{1 - q^ *  }}} } \left( {\frac{{d\tau _n }}{{d\lambda }}} \right)^2 } \right. \\
  + \left. {\tilde Z\left( \tau  \right)^{ - 1} \sum\limits_n {\left[ {1 + \left( {1 - q^ *  } \right)\tau _n } \right]^{\frac{{2 - q^ *  }}{{1 - q^ *  }}} \frac{{d^2 \tau _n }}{{d\lambda ^2 }}} } \right\}, \\
 \end{array}
\end{equation}

and

\begin{equation}
\begin{array}{l}
Term - 2 =  - \frac{1}{{\beta ^ *  }}\left\{ {\sum\limits_n {\left[ {1 + \left( {1 - q^ *  } \right)\tau _n } \right]^{\frac{1}{{1 - q^ *  }}} \frac{{d\tau _n }}{{d\lambda }}\left[ {D_{(q^ *)  }^\tau  \tilde Z\left( \tau  \right)^{ - 1} } \right]} } \right\}\frac{{d\tau }}{{d\lambda }} \\
  = \frac{1}{{\beta ^ *  }}\left\{ {\tilde Z\left( \tau  \right)^{ - 2} \sum\limits_n {\left[ {1 + \left( {1 - q^ *  } \right)\tau _n } \right]^{\frac{2}{{1 - q^ *  }}} \left( {\frac{{d\tau _n }}{{d\lambda }}} \right)^2 } } \right\}. \\
\end{array}
\end{equation}

\begin{equation}
\begin{array}{l}
 \Im _{q^ *  }^{\left( 2 \right)}  =  - \frac{1}{{\beta ^ *  }}\left\{ {\tilde Z\left( \tau  \right)^{ - 1} \sum\limits_n {\left[ {1 + \left( {1 - q^ *  } \right)\tau _n } \right]^{\frac{1}{{1 - q^ *  }}} \left( {\frac{{d\tau _n }}{{d\lambda }}} \right)} } \right.^2  \\
 \left. { + \tilde Z\left( \tau  \right)^{ - 1} \sum\limits_n {\left[ {1 + \left( {1 - q^ *  } \right)\tau _n } \right]^{\frac{{2 - q^ *  }}{{1 - q^ *  }}} \frac{{d^2 \tau _n }}{{d\lambda ^2 }}} } \right\} \\
  + \frac{1}{{\beta ^ *  }}\left\{ {\tilde Z\left( \tau  \right)^{ - 2} \sum\limits_n {\left[ {1 + \left( {1 - q^ *  } \right)\tau _n } \right]^{\frac{2}{{1 - q^ *  }}} \left( {\frac{{d\tau _n }}{{d\lambda }}} \right)^2 } } \right\}. \\
\end{array}
\end{equation}

Setting $ \tau_n=-\beta ^ *E_n $, (40) yields
\begin{equation}
\begin{array}{l}
 \Im _{q^ *  }^{\left( 2 \right)} =  \beta ^ *  \left\{ {\tilde Z\left( {\beta ^ *  } \right)^{ - 2} \sum\limits_n {\left[ {1 - \left( {1 - q^ *  } \right)\beta ^ *  E_n } \right]^{\frac{2}{{1 - q^ *  }}} \left( {\frac{{dE_n }}{{d\lambda }}} \right)^2 } } \right. \\
 \left. { - \tilde Z\left( {\beta ^ *  } \right)^{ - 1} \sum\limits_n {\left[ {1 - \left( {1 - q^ *  } \right)\beta ^ *  E_n } \right]^{\frac{1}{{1 - q^ *  }}} \left( {\frac{{dE_n }}{{d\lambda }}} \right)^2 } }
 \right\}\\
 + \tilde Z\left( {\beta ^ *  } \right)^{ - 1} \sum\limits_n {\left[ {1 - \left( {1 - q^ *  } \right)\beta ^ *  E_n } \right]} ^{\frac{{2 - q^ *  }}{{1 - q^ *  }}} \frac{{d^2 E_n }}{{d\lambda ^2 }}.\\
\end{array}
\end{equation}

With the aid of the Hellmann-Feynman theorem (4), the second-order
perturbation term in (28) is

\begin{equation}
\begin{array}{l}
\left. {\Im _{q^ *  }^{\left( 2 \right)} } \right|_{\lambda  = 0} =
\beta ^ *  \left\{ {\left( {\left\langle {H_{1 } } \right\rangle
_{q^ *  }^{(0)} } \right)^2 - \sum\limits_n {p_n\left( {E_n^{(0)} }
\right)\left( {H_{1_{nn} } } \right)} ^2 } \right\}\\
+ \left. {\tilde Z\left( {\beta ^ *  } \right)^{ 1-q^*}
\sum\limits_n {p_n \left( {E_n^{\left( 0 \right)} } \right)}
\frac{{\partial \left\langle n \right|H_1 \left| n \right\rangle
}}{{\partial
\lambda }}} \right|_{\lambda  = 0}.\\
\end{array}
\end{equation}

Employing (23) ($ \left. {\frac{{d \left| {n} \right\rangle
}}{{d\lambda  }}} \right|_{\lambda  = 0}  = \left| n \right\rangle
^{\left( 1 \right)} $) and (24), (42) acquires the compact form

\begin{equation}
\begin{array}{l}
\left. {\Im _{q^ *  }^{\left( 2 \right)} } \right|_{\lambda  = 0} =
\beta ^ *  \left\{ {\left( {\left\langle {H_{1 } } \right\rangle
_{q^ *  }^{(0)} } \right)^2 - \sum\limits_n {p_n\left(
{E_n^{(0)} } \right)\left( {H_{1_{nn} } } \right)} ^2 } \right\}\\
+ 2\tilde Z\left( {\beta ^ *  } \right)^{1-q^*  } \sum\limits_n {p_n
\left( {E_n^{\left( 0 \right)} } \right)} \sum\limits_{m \ne n}
{\frac{{\left| {H_{1_{nm} } } \right|^2 }}{{E_n^{\left( 0 \right)} -
E_m^{\left( 0 \right)} }}},\\
\end{array}
\end{equation}
where: $ H_{1_{nm} }  = ^{\left( 0 \right)} \left\langle n
\right|H_1 \left| m \right\rangle ^{\left( 0 \right)} $.  Note that
(43) is the direct equivalent of Eq. (6) in Ref. [8] and Eq. (24) in
Ref. [10]. In the classical limit, the last term in (43) vanishes,
since the Hamiltonians $ H_0 $ and $ H_1 $ commute. In keeping with
the gist of previously cited works on GVPA's [8,10], this paper
employs the classical harmonic oscillator to examine the properties
and efficacy of the \textit{q-deformed} GVPA principle introduced
herein. Thus, the second-order perturbation term becomes
\begin{equation}
\begin{array}{l}
\left. {\Im _{q^ *  }^{\left( 2 \right)} } \right|_{\lambda  = 0} =
\beta ^ *  \left\{ {\left( {\left\langle {H_{1 } } \right\rangle
_{q^ *  }^{(0)} } \right)^2 - \sum\limits_n {p_n\left(
{E_n^{(0)} } \right)\left( {H_{1_{nn} } } \right)} ^2 } \right\}.\\
\end{array}
\end{equation}

Thus, for commuting Hamiltonians $ H_0 $ and $ H_1 $, the
perturbation expansion of the GFE is

\begin{equation}
\begin{array}{l}
 \Im _{q^ *  } \left( \lambda  \right) =
 - \frac{1}{{\tilde\beta   }}\frac{{\tilde Z_{\left( {q^ *  } \right),0} \left( {\beta ^ *  } \right)^{q^ *-1  }  - 1}}{{ q^ *-1  }}+ \lambda \left\langle {H_1 } \right\rangle _{q^ *  }^{(0)}  \\
  + \frac{{\lambda ^2 }}{{2!}}\beta ^ *  \left\{ {\left( {\left\langle {H_1 } \right\rangle _{q^ *  }^{(0)} } \right)^2  - \sum\limits_n {p_n \left( {E_n^{(0)} } \right)\left( {H_{1_{nn} } } \right)^2 } } \right\} + \theta \left( {\lambda ^3 } \right) + ... \\
 \end{array}
\end{equation}

The above analysis and (17) make tacit assumptions concerning the
commutation between the Newtonian derivative $ \frac{d}{{d\lambda }}
$ and the summation sign.  To analyze the contribution of the cut-off in the \textit{q-deformed} GVPA model for $\beta^* > 0 $ and $E_n \ge 0$, parallels are drawn with the analysis in Ref. [8].  First, the summation sign $
\sum\limits_n {} $ is replaced by the integral $ \int {d\Gamma } $,
where: $ \Gamma = \prod\limits_s {\frac{{dx_s dp_s }}{h}} $,  $ p_s
$ is the canonical momentum, and $ h $ is Planck's constant. The
integration is performed over the phase-space region defined by: $
\left[ {1 - \left( {1 - q^ *  } \right)\beta ^ *  H} \right] \ge 0 \
$.  Defining: $ f = \left[ {1 - \left( {1 - q^ *  } \right)\beta ^
* H} \right]^{\frac{{1}}{1-q^*}} = \left[ {1 + \left( {1 - q^ *  } \right)\tau } \right]^{\frac{{1}}{1-q^*}}
=f(\tau)$, the following identity is employed by invoking (9) and
Theorem 1
\begin{equation}
\begin{array}{l}
\frac{d}{{d\lambda }}\int {d\Gamma } D_{\left( {q^ *  } \right)}^\tau  f\left( \tau  \right) = \int {d\Gamma } D_{\left( {q^ *  } \right)}^\tau  f\left( \tau  \right)\frac{{d\tau }}{{d\lambda }} \\
  + \int_{\partial V} {\sum\limits_u {dS_u } } \left( {D_{\left( {q^ *  } \right)}^{\tau _u } f\left( {\tau _u } \right)\frac{{d\tau _u }}{{dy_u }}\frac{{dy_u }}{{d\lambda }}} \right). \\
 \end{array}
\end{equation}

Note that the distinction of (46) \textit{vis-\'{a}-vis} Eq. (10) in
Ref. [8] is due to replacing: $ \frac{{df
\left( \tau \right)}}{{d\lambda }}$ by: $\delta _{\left( q^*
\right),\tau} f \left( \tau \right) $ (see (9)), in the
\textit{q-deformed} GVPA model. Here, $ f(\tau) $ is a function of
the phase-space variables $ y_u $, and, $
\partial V $ is the hypersurface defined by: $ \left[ {1 - \left( {1
- q^ * } \right)\beta ^ *  H} \right] = \left[ {1 + \left( {1
- q^ * } \right)\tau} \right]=0 $.  The $ n^{th} $-order
perturbation $\Im_{q^*}^{(n)}$ contains a term proportional to: $ \left[ {1 + \left( {1 - q^
*  } \right)\tau } \right]^{\frac{{n - \left( {n - 1} \right)q^ *
}}{{1 - q^ *  }}} $.  For example, $ \Im _{q^ *  }^{\left( 2
\right)} $ in (40), contains a term: $ \sum\limits_n {\left[ {1 +
\left( {1 - q^
* } \right)\tau _n } \right]^{\frac{{2 - q^ * }}{{1 - q^ *  }}} }
$ corresponding to $ n=2 $.  In order that $ \Im _{q^
* }^{\left( n \right)} $ \textit{not} contribute to the second term
in (46), the condition: $ n - \left( {n - 1} \right)q^ *   > 0
\Rightarrow q^* < 1 + \frac{1}{{n - 1}} $ is to be observed so that: $ D_{\left( {q^ *  } \right)}^\tau  f\left( \tau  \right) = 0 \Rightarrow f\left( \tau  \right) = 0 $ on $\partial V$.

\section{The generalized Bogoliubov inequality}

Within the context of the \textit{q-deformed} GVPA, the generalized
Bogoliubov inequality truncated to first-order terms is
\begin{equation}
\Im _{q^ *  }^{}  \le \Im _{q^ *  }^{\left( 0 \right)}  +
\left\langle {H_1 } \right\rangle _{q^ *  }^{\left( 0 \right)}.
\end{equation}

For a 1-D classical harmonic oscillator of mass $ M $ and angular
frequency $ \omega $, the Hamiltonian is:  $ H = \frac{{p^2 }}{{2M}}
+ \frac{{M\omega ^2 x^2 }}{2}$.  Here, $ x $ is the coordinate and $
p $ is the canonical momentum.  Following a procedure analogous to
that in Ref. [10], from (6), the canonical partition function is
expressed in continuous form as
\begin{equation}
\tilde Z\left( {\beta ^ *  } \right) = \int\limits_0^N {\left[ {1 -
\left( {1 - q^ *  } \right)\beta ^ *  n\delta _0 } \right]}
^{\frac{1}{{1 - q^ *  }}} dn = \frac{1}{{\left( {2 - q^ *  }
\right)\beta ^ *  \delta _0 }};\beta ^ *   > 0.
\end{equation}

Here, $ 0<q^*<2 $, $ N \to \infty $ for $q^*>1$, and $ N =
\frac{1}{{\left( {1 - q^ *  } \right)\beta ^ *  \delta _0 }} $ for
$q^*<1$. Akin to [10], $ \delta _0  = \hbar \omega $, is a positive
constant with units of energy. Also, $\hbar=h/{2\pi} $, where $h$ is
Planck's constant.  Thus, (26) is re-written as
\begin{equation}
\begin{array}{l}
 \Im _{q^ *  }  =  - \frac{1}{{\left( {q^ *-1 } \right)\tilde\beta }}\left[ {\left( {\frac{1}{{\left( {2 - q^ *  } \right)\beta ^ *  \delta _0 }}} \right)^{q^ * -1 }  - 1} \right]. \\
 \end{array}
\end{equation}

The unperturbed Hamiltonian for a particle of mass $M$ in a 1-D box
is: $ H_0  = \frac{{p^2 }}{{2M}} + V_0 $, where $V_0=0$ for $|x|<
L/2$ and $V_0 \rightarrow \infty$ for $|x|\ge L/2$.  In accordance with [10],
the continuous energy spectrum of the particle is: $ E_n^{\left( 0
\right)}  = \frac{{\delta _b^2 n^2 }}{{2ML^2 }} $, where $\delta_b=
\hbar \pi$ is a constant with the dimension of action.  Thus, (29)
is described in terms of the Euler $\Gamma$ function as
\begin{equation}
\begin{array}{l}
\tilde Z_{\left( {q^ *  } \right),0} \left( {\beta ^ *  } \right) =
\int\limits_0^N {\left[ {1 - \left( {1 - q^ *  } \right)\beta ^ *
\frac{{n^2 \delta _b^2 }}{{2ML^2 }}} \right]^{\frac{1}{{1 - q^ * }}}
dn}  \\
= \left. \begin{array}{l}
 \frac{L}{{\delta _b^{} }}\sqrt {\frac{{M\pi }}{{2\left( {1 - q^ *  } \right)\beta ^ *  }}} \frac{{\Gamma \left( {\frac{{2 - q^ *  }}{{1 - q^ *  }}} \right)}}{{\Gamma \left( {\frac{3}{2} + \frac{1}{{1 - q^ *  }}} \right)}};q^ *   < 1, \\
 \frac{L}{{\delta _b^{} }}\sqrt {\frac{{M\pi }}{{2\left( {q^ *   - 1} \right)\beta ^ *  }}} \frac{{\Gamma \left( { - \frac{1}{2} + \frac{1}{{q^ *   - 1}}} \right)}}{{\Gamma \left( {\frac{1}{{q^ *   - 1}}} \right)}};3 > q^ *   > 1 \\
 \end{array} \right\},
\end{array}
\end{equation}
where $N \rightarrow \infty$ for $ q^*>1$ and $ N = \frac{L}{{\delta
_b^{} }}\sqrt {\frac{{2M}}{{\left( {1 - q^ *  } \right)\beta ^ * }}}
$ for $q^*<1$.\footnote{Symbolic integration was performed using
$MATHEMATICA^{\circledR} $.} Note that (50) is identical to Eq. (11)
in [8], with $q^* $ replacing $q$.

Following the procedure employed in [10], the matrix elements of $
H_{1_{nm} } $ in (33) and (43) are (Eq. (33) in [10])
\begin{equation}
H_{1_{nm} }  = \delta _{nm} 2\int\limits_{ - L/2}^{L/2}
{\frac{{M\omega ^2 x^2 }}{2}} \frac{{dx}}{{2L}} = \delta _{nm}
\frac{{M\omega ^2 L^2 }}{{24}}.
\end{equation}

Here, $\delta_{nm} $ is the Kronecker delta.  Employing (50),
the first-order perturbation term (33) becomes
\begin{equation}
\begin{array}{l}
 \left. {\Im _{q^ *  }^{\left( 1 \right)} } \right|_{\lambda  = 0}  = \left\langle {H_1 } \right\rangle _{q^ *  }^{\left( 0 \right)}  = \sum\limits_n {p_n \left( {E_n^{\left( 0 \right)} } \right)H_{1_{nn} } }  \\
  = \frac{{M\omega ^2 L^2 }}{{24}}\frac{1}{{\tilde Z_{\left( {q^ *  } \right),0} \left( {\beta ^ *  } \right)}}\int\limits_0^N {\left[ {1 - \left( {1 - q^ *  } \right)\beta ^ *  \frac{{n^2 \delta _b^2 }}{{2ML^2 }}} \right]} ^{\frac{1}{{1 - q^ *  }}} dn = \frac{{M\omega ^2 L^2 }}{{24}}. \\
 \end{array}
\end{equation}

From (25),(26), (29), (48)-(50), and (52), the  generalized
Bogoliubov inequality (47) is
\begin{equation}
\begin{array}{l}
 \Im _{q^ *  }^{}  \le \Im _{q^ *  }^{\left( 0 \right)}  + \left\langle {H_1 } \right\rangle _{q^ *  }^{\left( 0 \right)}  \\
  \Rightarrow  - \frac{1}{{\left( {q^ *-1  } \right)\tilde\beta }}\left[ {\tilde
Z \left( {\beta ^ *  } \right)^{(q^ *-1) }  - 1} \right] =  - \frac{1}{{\left( {q^ *-1  } \right)\tilde\beta }}\left[ {\left( {\frac{{2\pi }}{{\left( {2 - q^ *  } \right)\beta ^ *  h\omega }}} \right)^{\left( {q^ *-1  } \right)}  - 1} \right] \\
\le  - \frac{1}{{\left( {q^ *-1  } \right)\tilde\beta }}\left[ {\tilde Z_{\left( {q^ *  } \right),0} \left( {\beta ^ *  } \right)^{\left( {q^ *-1  } \right)}  - 1} \right] + \frac{{M\omega ^2 L^2 }}{{24}}, \\
 where \\
 \tilde Z_{\left( {q^ *  } \right),0} \left( {\beta ^ *  } \right) = \left. \begin{array}{l}
 \frac{L}{h}\sqrt {\frac{{2M\pi }}{{\left( {1-q^ *  } \right)\beta ^ *  }}} \frac{{\Gamma \left( {\frac{{2 - q^ *  }}{{1 - q^ *  }}} \right)}}{{\Gamma \left( {\frac{3}{2} + \frac{1}{{1 - q^ *  }}} \right)}};q^ *   < 1 \\
 \frac{L}{h}\sqrt {\frac{{2M\pi }}{{\left( {q^ *   - 1} \right)\beta ^ *  }}} \frac{{\Gamma \left( { - \frac{1}{2} + \frac{1}{{q^ *   - 1}}} \right)}}{{\Gamma \left( {\frac{1}{{q^ *   - 1}}} \right)}};3 > q^ *   > 1 \\
 \end{array} \right\}. \\
 and \\
 \tilde\beta  = \beta/q=\beta ^ *  \tilde Z\left( {\beta ^ *  } \right)^{q^ *   - 1}  = \beta ^ *  \left( {\frac{{2\pi }}{{\left( {2 - q^ *  } \right)\beta ^ *  h\omega }}} \right)^{q^ *   - 1} . \\
 \end{array}
\end{equation}

The value of $L$ is obtained by minimizing the right hand side of
(53), yielding
\begin{equation}
\begin{array}{l}
  - \frac{{L^{  \left( { q^ *-3  } \right)} }}{\tilde\beta }C_{q^ *  } \left( {\beta ^ *  } \right) + \frac{{M\omega ^2 }}{{12}} = 0 \Rightarrow L = \left[ {\frac{{\tilde\beta M\omega ^2 }}{{12C_{q^ *  } \left( {\beta ^ *  } \right)}}} \right]^{  \frac{1}{{q^ *-3  }}} , \\
 where \\
 C_{q^ *  } \left( {\beta ^ *  } \right) = \left. \begin{array}{l}
 \left( {\sqrt {\frac{{2M\pi }}{{h^2 \left( {1 - q^ *  } \right)\beta ^ *  }}} \frac{{\Gamma \left( {\frac{{2 - q^ *  }}{{1 - q^ *  }}} \right)}}{{\Gamma \left( {\frac{3}{2} + \frac{1}{{1 - q^ *  }}} \right)}}} \right)^{q^ *-1  } ;q^ *   < 1 \\
 \left( {\sqrt {\frac{{2M\pi }}{{h^2 \left( {q^ *   - 1} \right)\beta ^ *  }}} \frac{{\Gamma \left( { - \frac{1}{2} + \frac{1}{{q^ *   - 1}}} \right)}}{{\Gamma \left( {\frac{1}{{q^ *   - 1}}} \right)}}} \right)^{q^ *-1  } ;3 > q^ *   > 1 \\
 \end{array} \right\}. \\
 \end{array}
\end{equation}

Here, (54) demonstrates that $L$ accounts for both
\textit{sub-additivity} ($q^*>1$) and \textit{super-additivity}
($q^*<1$). This feature is absent in formulations of the generalized
Bogoliubov inequality truncated at first-order terms in Refs.
[8,10].\footnote{Note that \textit{sub-additivity} and
\textit{super-additivity} are defined in terms of $q^*$ because (53)
and (54) are parameterized by $q^*$.} Specifically, Eq. (13) in Ref.
[8] and Eq. (40) in Ref. [10] make no allowance for
\textit{sub-additivity} and \textit{super-additivity} in the
expression for $L$.

\section{Numerical studies}
The \textit{q-deformed} GVPA model is numerically studied by
evaluating the generalized Bogoliubov inequality (53) for values of
$ \beta^* \in [0.01,3.5] $, akin to the \textit{parametric
perspective} described in [18] for various values of $q^*$.  Here,
$M=\omega=h=1$. Representative examples for the generalized
Bogoliubov inequality (47) and (53) are demonstrated in Figure 1,
for $q^*=0.5 $ and $q^*=0.95 $. Numerical examples for the
generalized Bogoliubov inequality for $q^*=1.3 $ and $q^*=1.75$ are
depicted in Figure 2.

From Figure 1 and Figure 2, it is readily observed that the
generalized Bogoliubov inequality in (47) and (53) tends to an
equality with decreasing values of $q^*$.   Specifically, the
difference between the exact solution ($\Im_{q^*}$) and the
perturbation solution ($\Im_{q^*}^{(0)}+ <H_1>_{q^*}^{(0)}$)
decreases with decreasing $q^*$.  Note that the exact solution of
the GFE (the LHS of the inequality in (47) and (53)) almost exactly
coincides with the perturbation solution, for $q^*=0.5 $. The
relations between $\tilde\beta$ and $\beta^*$ are displayed in
Figure 3 for both $q^*<1$ and $q^*>1$ .

The expression for $L$ (54) is the most explicit manifestation of
\textit{sub-additivity} and \textit{super-additivity} possessed by
the \textit{q-deformed} GVPA, for $k^{th}$-order perturbation
expansions of the $q^*-deformed$ GFE (26), $k\ge 1$. Figure 4
displays the dependence of L on $\beta^*$.  For $0<q^*<1$, L
\textit{increases} with increasing values of $q^*$.  In contrast,
for $1<q^*<3$, L \textit{decreases} with increasing values of $q^*$.

\section{Summary and conclusions}

The theoretical framework for a \textit{q-deformed} GVPA model which
generalizes previous works on GVPA models [8-10], has been formulated.  The underlying theory for the variational procedure of the \textit{q-deformed} GVPA model employs \textit{q-deformed} calculus [21].  The significant feature of the \textit{q-deformed} GVPA is that
expectation values may be self-consistently formulated in the form
of normal averages, instead of the Curado-Tsallis form
[8-10]. This feature acquires special significance owing to recent results
that establish that expectations in the normal averages form, in contrast to the vastly more utilized $q$-averages form, are physical and consistent with both the generalized H-theorem and the generalized \textit{Stosszahlansatz} (molecular chaos hypothesis) [19, 20].

It is qualitatively and quantitatively demonstrated that the
\textit{q-deformed} GVPA model exhibits both \textit{sub-additivity}
and \textit{super-additivity} in terms of the nonadditivity
parameter $q^* $ for the generalized Bogoliubov inequality,
truncated at first-order terms. This property is not possessed by
previous GVPA models [8, 10].  Specifically, it may be construed
that the \textit{q-deformed} GVPA presented in this paper exhibits
authentic characteristics of a VPA even for the generalized
Bogoliubov inequality truncated at first-order terms.  In contrast,
previous cited analyses [8, 10] do not exhibit any equivalent
property, rendering them purely variational principles when the
generalized Bogoliubov inequality is truncated at first-order terms.
Numerical simulations that demonstrate the results and the efficacy
of the \textit{q-deformed} GVPA are presented.

Future works that will be presented elsewhere accomplish a
three-fold objective: $(i)$ a comparative study taking into account higher-order terms between the
\textit{q-deformed} GVPA model presented in this paper, its counterpart based on the \textit{dual Tsallis entropy}: $S_{q^*=2-q} $ (see Ref. [3] and the references therein), and the results of previous cited
studies [8-10], $(ii)$ a
\textit{q-deformed} GVPA model for the homogeneous Arimoto entropy
[23] with normal averages constraints, defined in terms of the
\textit{escort probability} [24]. The homogeneous nonadditive
Arimoto entropy is defined in terms of the \textit{escort
probability} as: $ S_q^H \left( P \right) =
 - \frac{{\left( {\sum\limits_i {P_i^{\frac{1}{q}} } } \right)^q  - 1}}{{q - 1}} $, and,
$(iii)$ use of the \textit{q-deformed} GVPA to analyze critical point behavior in deterministic
annealing [25], within the framework of generalized statistics, and,
the deformed statistics information bottleneck method [4] in machine
learning.

\textbf{Acknowledgements}

RCV gratefully acknowledges support from \textit{RAND-MSR} contract
\textit{CSM-DI $ \ \& $ S-QIT-101155-03-2009}.  RCV gratefully acknowledges discussions with E. P. Borges concerning the \textit{q-deformed} generalization of the variational-perturbation approximation.  The authors express their gratitude to the anonymous reviewers for their constructive comments.

\newpage

\section*{FIGURE CAPTIONS}

\textbf{Fig. 1:}  Generalized Bogoliubov inequality for the
$q-deformed$ GVPA model for $q^*=0.5$ and $q^*=0.95$.  Here,
$\Im_{q^*}$ is the LHS of (53) (the exact solution) and
$\Im_{q^*}^{(0)}+<H_1>_{q^*}^{(0)}$ is the RHS of (53) (the
perturbation solution). The generalized Bogoliubov inequality
increases with increasing $q^*$.  Note the near overlap of the exact
solution and the perturbation solution for $q^*=0.5$.
\\
\textbf{Fig. 2:}  Generalized Bogoliubov inequality for the
$q-deformed$ GVPA model for $q^*=1.3 $ and $q^*=1.75$.  Here,
$\Im_{q^*}$ is the LHS of (53) (the exact solution) and
$\Im_{q^*}^{(0)}+<H_1>_{q^*}^{(0)}$ is the RHS of (53) (the
perturbation solution).  Note the increase in the generalized Bogoliubov inequality with increasing $q^*$.
\\
\textbf{Fig. 3:}  Relation between the scaled "inverse thermodynamic
temperature" ($\beta/q=\tilde\beta $) and the "physical inverse temperature" ($\beta^*$) for $q^*$=0.5, 0.95, 1.3, and, 1.75.
\\
\textbf{Fig. 4:}  Dependence of $L$ on $\beta^*$. Note the increase in the value of $L$ for increasing $q^*<1$, and the decrease in the value of $L$ for increasing $q^*>1$.
\newpage

\begin{figure}[thpb]
\centering
\begin{center}
\includegraphics[scale=1.00]{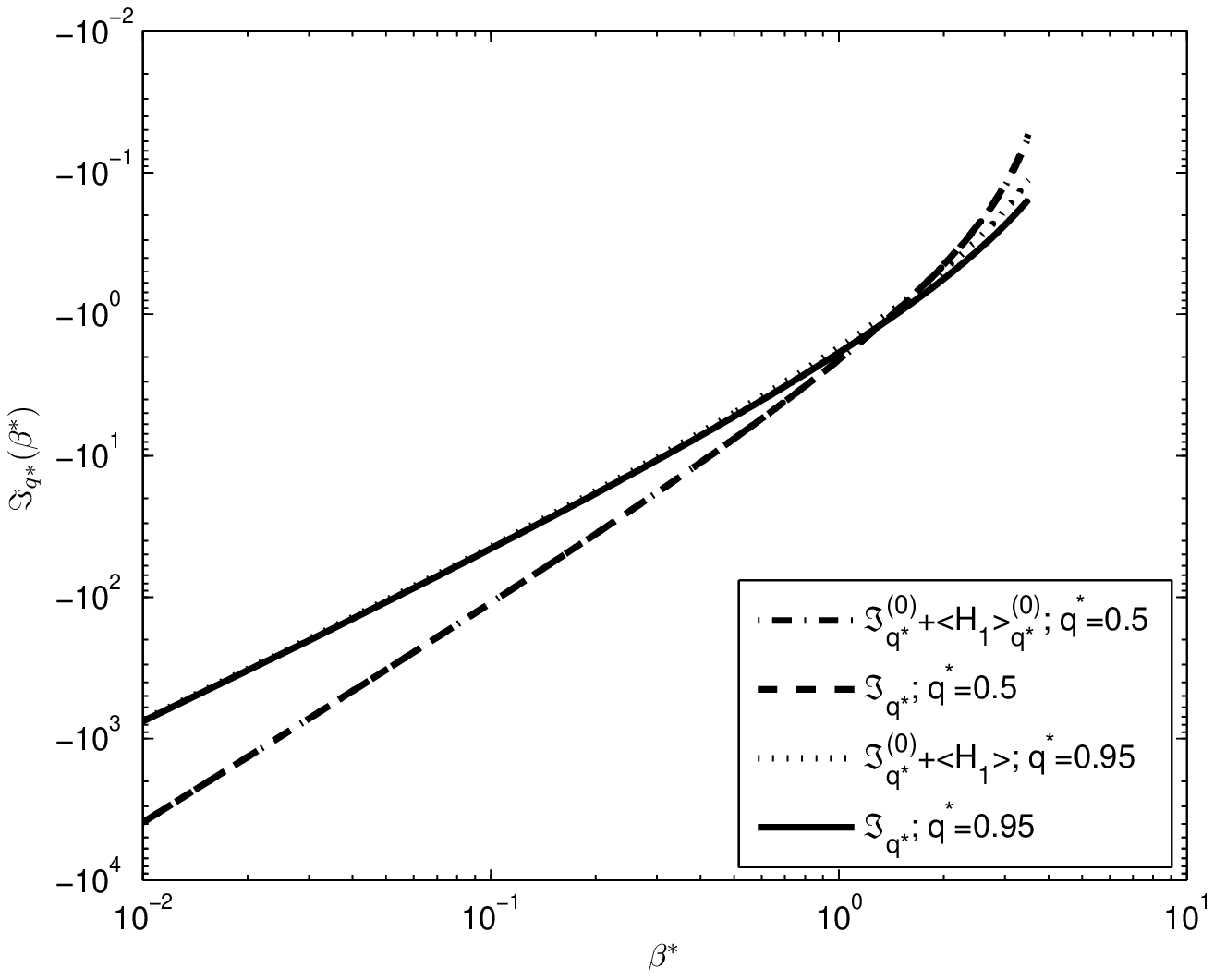}
\end{center}
\end{figure}

\newpage

\begin{figure}[thpb]
\centering
\begin{center}
\includegraphics[scale=1.00]{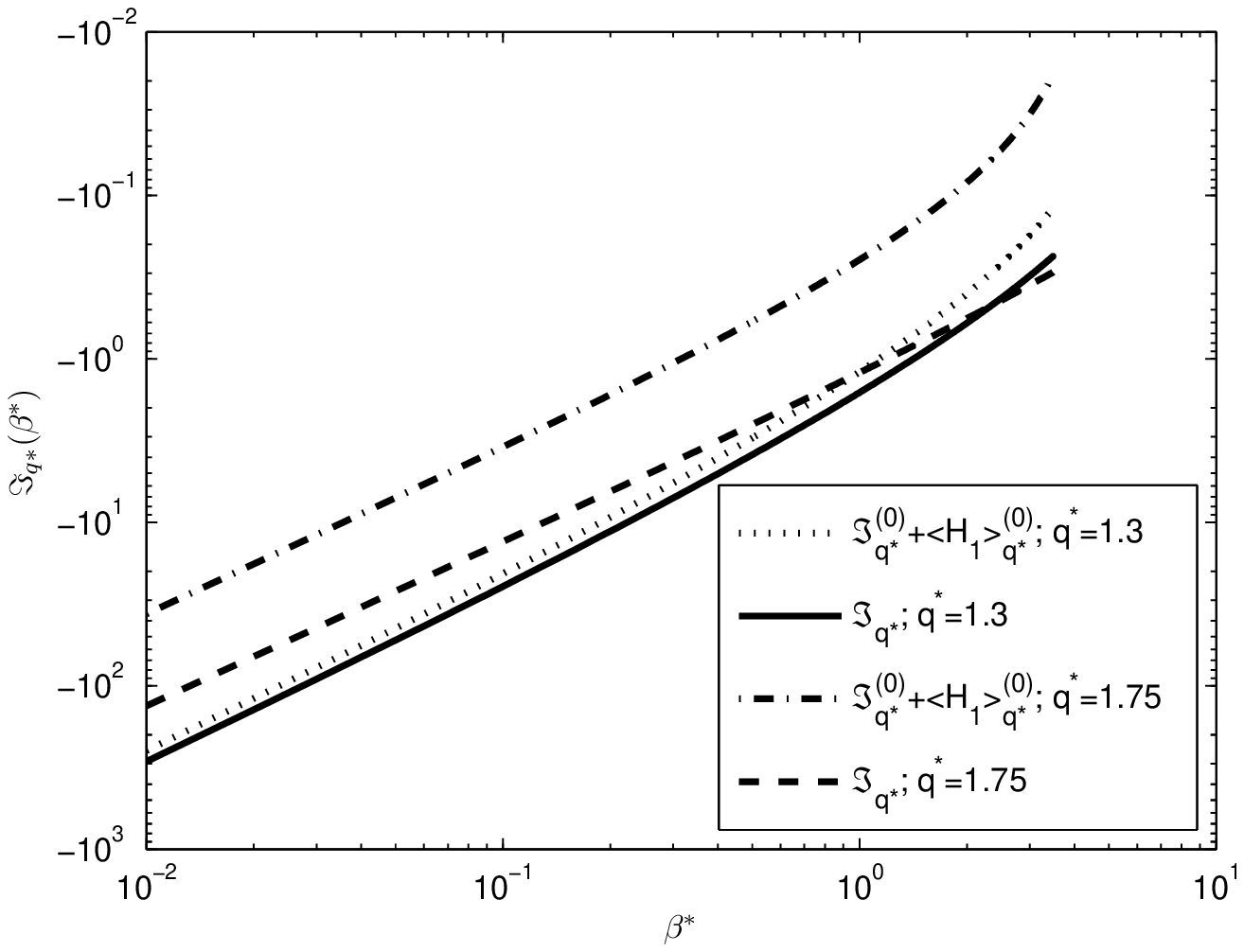}
\end{center}
\end{figure}

\newpage

\begin{figure}[thpb]
\centering
\begin{center}
\includegraphics[scale=1.00]{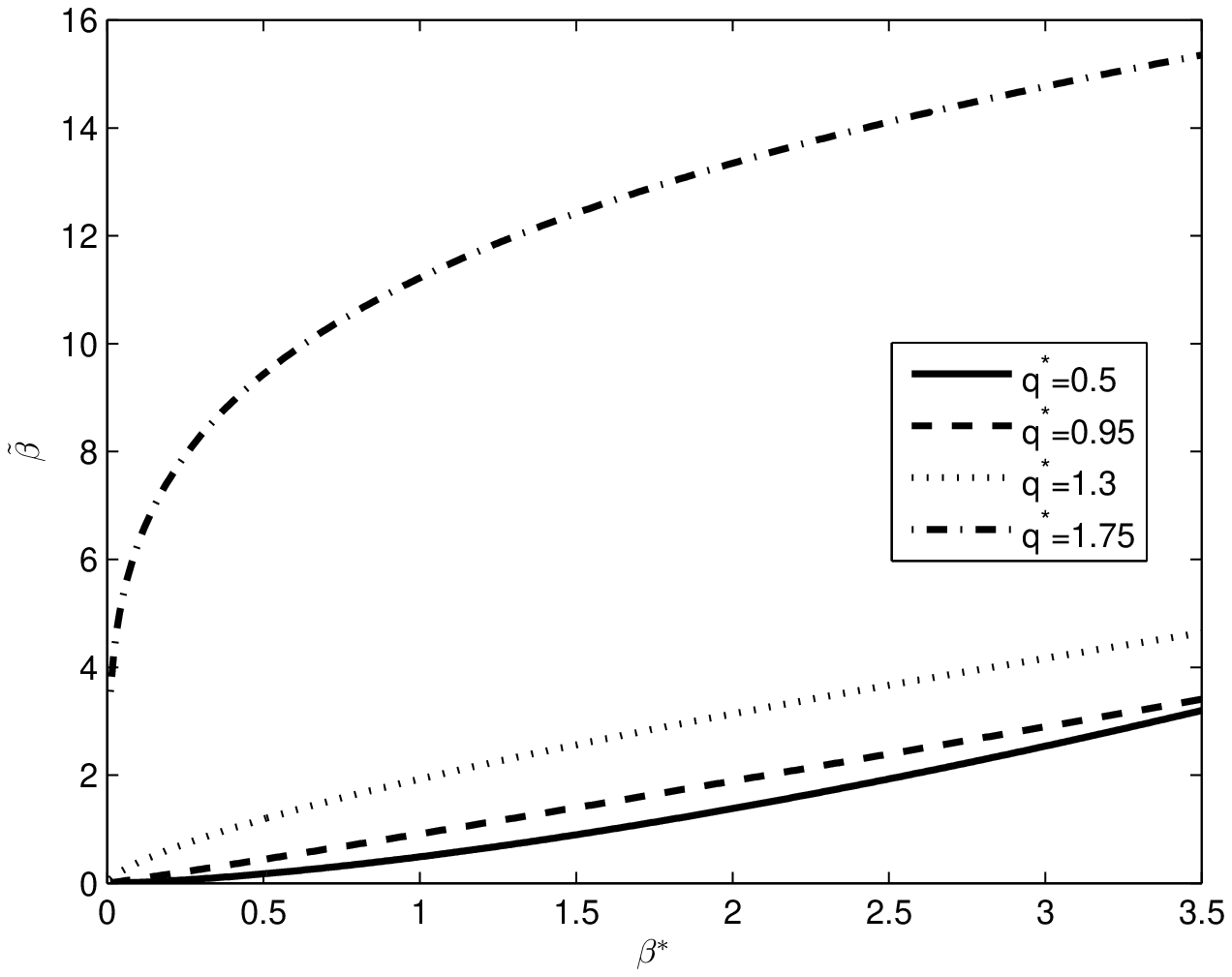}
\end{center}
\end{figure}

\newpage

\begin{figure}[thpb]
\centering
\begin{center}
\includegraphics[scale=1.00]{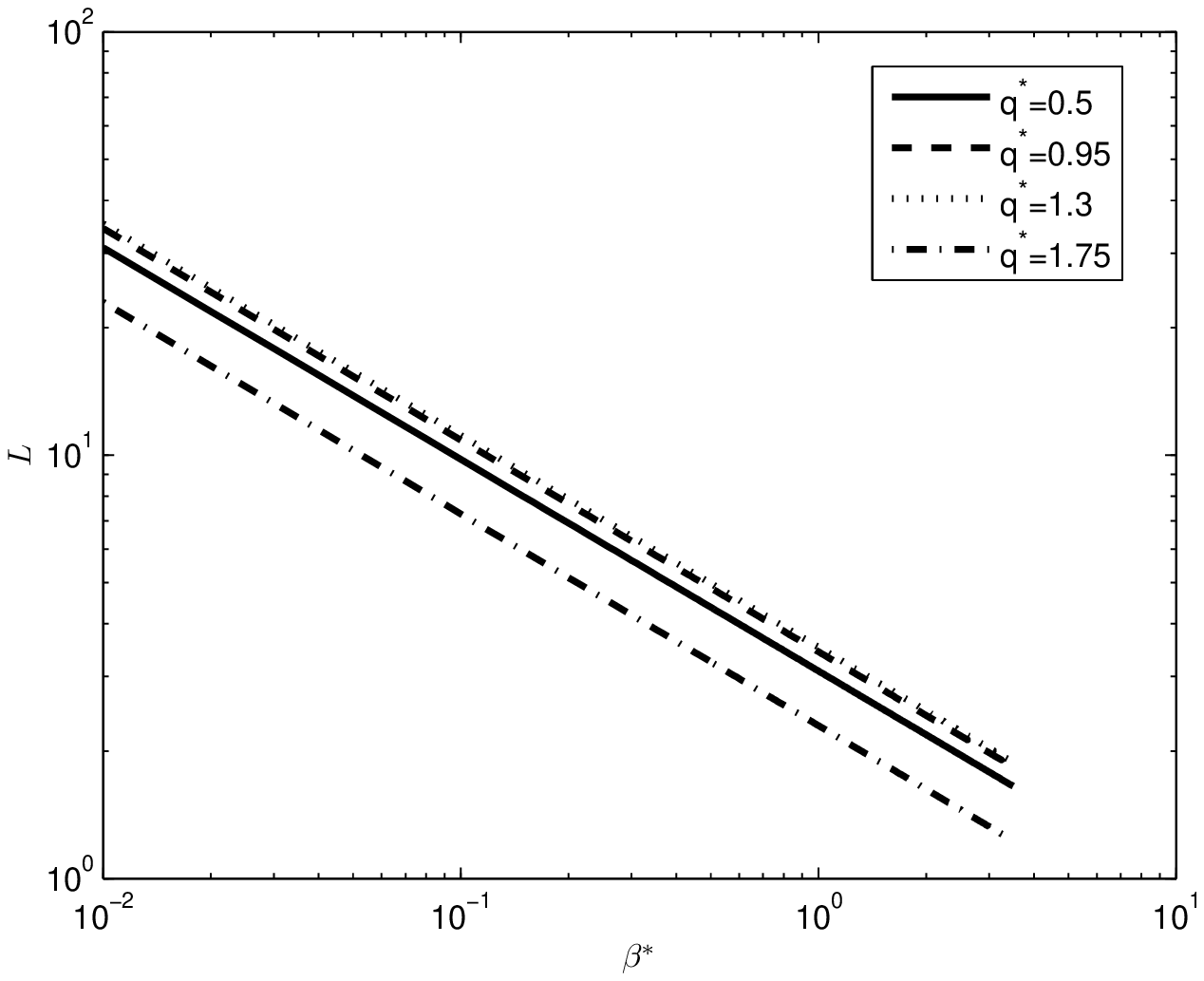}
\end{center}
\end{figure}

\newpage

\appendix
\numberwithin{equation}{section}
\renewcommand{\theequation}{A.\arabic{equation}}      

\section*{Appendix A:  Derivation of expression for $q^*$-deformed
Generalized Free Energy}

The canonical probability that maximizes the Tsallis entropy using
constraints defined in terms of normal averages is (Eq. (23) in Ref. [18])
\begin{equation}
\begin{array}{l}
 p_n  = \left[ {\aleph _q  + \frac{{\left( {q - 1} \right)}}{q}\beta U_q  - \frac{{\left( {q - 1} \right)}}{q}\beta E_n } \right]^{\frac{1}{{q - 1}}}  = \frac{{\left[ {1 - \frac{{\left( {q - 1} \right)}}{{\aleph _q  + \left( {q - 1} \right)\tilde \beta U_q }}\tilde \beta E_n } \right]^{\frac{1}{{q - 1}}} }}{{\tilde Z\left( {\tilde \beta } \right)}}; \\
 where \\
 U_q  = \sum\limits_n {p_n E_n } ,\tilde \beta  = \frac{\beta }{q},\aleph _q  = \sum\limits_n {p_n^q } ,and,\tilde Z\left( {\tilde \beta } \right) = \left( {\aleph _q  + \left( {q - 1} \right)\tilde \beta U_q } \right)^{\frac{1}{{1 - q}}}.  \\
 \end{array}
\end{equation}

From the definition of the partition function in (A.1), the
following thermodynamic relation is obtained
\begin{equation}
\ln _q \tilde Z\left( {\tilde \beta } \right) = \frac{{\aleph _q  +
\left( {q - 1} \right)\tilde \beta U_q  - 1}}{{1 - q}} \Rightarrow
\frac{{d\ln _q \tilde Z\left( {\tilde \beta } \right)}}{{d\tilde
\beta }} =  - U_q.
\end{equation}

The Tsallis entropy is

\begin{equation}
S_q  = \frac{{\aleph _q  - 1}}{{1 - q}}.
\end{equation}

Using the definition of $\tilde Z(\tilde\beta)$ in (A.1), (A.3)
yields the thermodynamic relation
\begin{equation}
S_q  = \frac{{\tilde Z\left( {\tilde \beta } \right)^{1 - q}  +
\left( {1 - q} \right)\tilde \beta U_q  - 1}}{{1 - q}} \Rightarrow
\frac{{dS_q }}{{dU_q }} = \tilde \beta.
\end{equation}

From (A.4), the GFE is thus defined as
\begin{equation}
\begin{array}{l}
 F_q = U_q  - \frac{1}{{\tilde \beta }}S_q  \\
  = U_q  - \frac{1}{{\tilde \beta }}\left( {\frac{{\tilde Z\left( {\tilde \beta } \right)^{1 - q}  + \left( {1 - q} \right)\tilde \beta U_q  - 1}}{{1 - q}}} \right) =  - \frac{1}{{\tilde \beta }}\ln _q \tilde Z\left( {\tilde \beta } \right). \\
 \end{array}
\end{equation}

From (A.1) and (A.5)

\begin{equation}
\begin{array}{l}
 F_q  =  - \frac{1}{{\tilde \beta }}\ln _q \sum\limits_n {\left[ {1 - \left( {q - 1} \right)\frac{{\tilde \beta }}{{\tilde Z\left( {\tilde \beta } \right)^{1 - q} }}E_n } \right]} ^{\frac{1}{{q - 1}}}  \\
  =  - \frac{1}{{\tilde \beta }}\ln _q \sum\limits_n {\left[ {1 - \left( {q - 1} \right)\beta ^ *  E_n } \right]} ^{\frac{1}{{q - 1}}}  =  - \frac{1}{{\tilde \beta }}\ln _q \tilde Z\left( {\beta ^ *  } \right), \\
 \end{array}
\end{equation}
where: $\beta ^ *   = \frac{{\tilde \beta }}{{\tilde Z\left( {\tilde \beta } \right)^{1 - q} }}$.  Here, $ \tilde Z\left( {\beta ^ *  } \right) = \sum\limits_n {\left[ {1 - \left( {q - 1} \right)\beta ^ *  E_n } \right]} ^{\frac{1}{{q - 1}}}$.

Invoking the \textit{additive duality}: $ q^*=2-q$ (see Ref. [1] and
the references therein) resulting in: $\tilde Z(\beta^*)= \sum\limits_n {\left[ {1 - \left( {1-q^*} \right)\beta ^ *  E_n } \right]} ^{\frac{1}{{1-q^*}}}$, setting: $\tau_n=-\beta^*E_n$  and employing Eq. (25) of this paper, the \textit{$q^*$-deformed} GFE (Eq. (26) of this paper) is expressed as
\begin{equation}
F_{q \to q^ *   = 2 - q}  = \Im _{q^ *  }  =  - \frac{1}{{\tilde
\beta }}\frac{{\tilde Z\left( {\beta ^ *  } \right)^{q^ *   - 1}  -
1}}{{q^ *   - 1}}= - \frac{1}{{\tilde \beta }}\frac{{\tilde Z\left(
\tau  \right)^{q^ *   - 1}  - 1}}{{q^ *   - 1}}.
\end{equation}

\end{document}